\documentclass[12pt,preprint]{aastex}

\lefthead{Pippo}

\righthead{The double Subgiant Branch of NGC~1851}

\begin{document}
\title{The double Subgiant Branch of NGC~1851: the role of the CNO abundance}

\author{S. Cassisi\altaffilmark{1}, M. Salaris\altaffilmark{2,3}, A. Pietrinferni\altaffilmark{1}, G. Piotto\altaffilmark{4}, 
A.P. Milone\altaffilmark{4}, L.R. Bedin\altaffilmark{5}, J. Anderson\altaffilmark{6}}

\altaffiltext{1}{INAF - Osservatorio Astronomico di Collurania, Via M.\ Maggini,
I-64100 Teramo, Italy; cassisi,pietrinferni@oa-teramo.inaf.it}

\altaffiltext{2}{Astrophysics Research Institute, Liverpool John Moores University,
Twelve Quays House, Egerton Wharf, Birkenhead CH41 1LD, UK;ms@astro.livjm.ac.uk}

\altaffiltext{3}{Max-Planck-Institut f\"ur Astrophysik, Karl-Schwarzschild-Strasse~1, Garching D-85748, Germany}

\altaffiltext{4}{Dipartimento di Astronomia, Universit\'a di Padova, Vicolo dell'Osservatorio 3, I-35122 Padua, Italy; giampaolo.piotto@unipd.it}

\altaffiltext{5}{Space telescope Science Institute, 3700 San Martin Drive, Baltimore, MD 21218, USA}

\altaffiltext{6}{Department of Physics and Astronomy, Mail Stop 108, Rice University, 6100 Main Street, Houston, TX 77005; jay@eeyore.rice.edu}

\begin{abstract}
We explore the possibility that the anomalous split in the Subgiant branch of the galactic globular cluster NGC~1851 is 
due to the presence of two distinct stellar populations with very different 
initial metal mixtures: a normal $\alpha-$enhanced component, and one characterized by strong anticorrelations 
among the CNONa abundances, with a total CNO abundance increased by a factor of two. 
We test this hypothesis taking into account various empirical constraints, and conclude that the two populations 
should be approximately coeval, with the same initial He-content. More high-resolution spectroscopical measurements of heavy 
elements -- and in particular of the CNO sum -- for this cluster are necessary to prove (or disprove) this scenario.
\end{abstract}


\section{Introduction}

Recent accurate spectroscopic and photometric observations of stars in Galactic Globular Clusters (GCs) have 
shown that surface
abundance variations of C, N, O, Na, and often also of Mg and Al, exist in stars within individual 
clusters (see, e.g., Gratton, Sneden \& Carretta~2004), and
that these elements display a pattern of abundance variations that is constant along the Red Giant Branch
(RGB) and down to the Turn-Off (TO) see, e.g., Gratton et al.~2001, Carretta et al.~2005).
This pattern shows anticorrelations
between CN and ONa (and, when observed, MgAl) overimposed onto a 'normal' $\alpha$-enhanced
heavy element distribution ([$\alpha$/Fe]$\sim 0.3-0.4$), in the sense that negative variations of C and O
are accompanied by increased N and Na abundances.
The generally accepted explanation is that these stars were born with the
observed CNONa abundance patterns. Winds of intermediate-mass Asymptotic Giant Branch (AGB) stars or winds
of fast rotating massive stars have been
invoked as sources of the necessary heavy element pollution (see, e.g. Ventura et
al.~2001, Decressin et al.~2007, D'Antona \& Ventura~2007 and references therein).
Provided that a significant fraction of this material is not
lost from the GC, new stars can form directly out of this matter (within
$10^6 - 10^8$ years after the formation of the first generation, the exact time depending on the nature
of the pollution sources) or from pre-existing matter polluted to varying degrees by the stellar ejecta.
The iron abundances would be constant among the different
subpopulations, as is actually observed in almost all GCs (see, e.g., Suntzeff~1993).

A prediction of both scenarios is that the second stellar generation
in a GC must have an enhanced initial He content, compared to the first generation with a 'normal'
$\alpha$-enhanced heavy element distribution. A spread in the initial He content of GC stars allows to interpret the
morphology of the Horizontal Branch (HB) in a number of GCs
(see, e.g., D'Antona \& Caloi~2004, 2007 and reference therein).
Recent accurate photometries of GCs obtained with the Advanced Camera for Survey (ACS)
on board of the $HST$ have given direct photometric evidence of multiple populations
with varying He content in $\omega$~Centauri (Anderson 1997, Bedin et al.~2004, Piotto et al.~2005). 
Given that this cluster 
has since long been known to harbor multiple populations with different [Fe/H] values, it may not be considered to be a typical GC,
rather a small galaxy. More remarkable, from the point of view of studying 'normal' GCs, is
the recent detection of a triple MS in NGC~2808 (Piotto et al.~2007). This is  a clear sign of populations with 
distinct initial He abundances (as early suggested by D'Antona et al.~2005) in a typical GC with 
no sign of spread in [Fe/H], hence no sign of metal enrichment due to supernovae ejecta.
An even more recent result is the discovery by Milone et al. (2007 -- hereafter M07) that the SGB of NGC~1851
is split into two distinct branches. This is, to date, the third GC for which there is direct
photometric evidence of multiple stellar populations.

The CMD by M07 displays a very narrow MS and RGB, that put
strong constraints on the possible interpretations of the SGB splitting. M07 discuss at length this issue
and conclude that two populations with the same age but with either a different initial He content or a different [Fe/H]
can be excluded. Another possibility they considered is to have the second population with both increased He ($Y \sim$0.30) and
higher [Fe/H] (by $\sim$ 0.2~dex). Given the opposite effect of metals and He on the colours of MS and RGB, this combination
would preserve the narrowness of MS and RGB and also reproduce the split of the SGB while keeping the age constant. The very recent 
results from high resolution spectroscopy 
of 8 bright cluster giants (an admittedly small sample) by Yong \& Grundahl~(2007) 
exclude the presence of a bimodal Fe abundance. The same conclusion can be reached by considering Walker~(1998) 
estimate of [Fe/H] from the Fourier decomposition of the light curves of 9 cluster RRab variables.
As an alternative scenario, M07 suggest that the SGB splitting could be simply due to the an age effect: the two SGBs
would correspond to two distinct star formation bursts separated by $\sim 1$Gyr.

Here we propose a connection between this photometric evidence of multiple populations in NGC~1851, and the pattern of CNONa
abundance anomalies typical of several GCs. This point has been raised by M07, but no specific calculations and/or quantitative estimates
have been made by the authors. 
We focus in particular on the role played by variations in the total CNO abundance, 
given the well known fact (see, e.g., Rood \& Crocker~1985, 
VandenBerg~1985, Salaris, Chieffi \& Straniero~1993 and references therein) that differences in the sum of the CNO nuclei affect 
both the TO and SGB level of isochrones of a fixed age and [Fe/H]. 
Theoretical models for the sources of heavy element pollution are presently unable to agree on quantitative 
predictions regarding the total CNO abundance in the polluting gas (and the extent of the CNONa anticorrelations) 
but some calculations suggest a possible increase of CNO in the second stellar generation. 
Ventura \& D'Antona~(2005a,b) analyzed in detail 
the various sources of uncertainties in case of AGB pollution, and with their best choice for the convection treatment 
they obtain CNO sums that can increase by a factor of about two (together with an enhanced He abundance) compared to the value of the 
first stellar generation. From the observational point of view, in case of NGC1851 
Hesser et al.~(1982) obtained 4\AA \ resolution spectra of bright stars
in the cluster. They found that three out of eight of their RGB stars show very strong CN bands, that
could be interpreted as due to the presence of carbon stars or to a very
large overabundance of nitrogen (or a combination of both carbon and nitrogen enhancements). Yong \& Grundahl~(2007) 
did not found a statistically significant  MgAl anticorrelation in their 8-star sample, but detected a clear ONa anticorrelation, 
whereas C and N abundances were not measured.
As for other GCs with spectroscopic measurements of the CNO sum, Cohen \& Melendez~(2005) found for example a 
a total CNO abundance constant within $\sim$0.12~dex in M13. On the other hand, in case of NGC6397, NGC6751 and 47~Tuc 
analyzed by Carretta et al.~(2005), error bars are such that a spread in CNO by a factor up to $\sim$2 cannot be excluded. 

As a working scenario we will assume the presence
of two distinct stellar populations in this cluster, one with a \lq{normal}\rq $\alpha$-enhanced metal distribution, 
and a second one with a
pattern of CNONa anticorrelations overimposed onto the normal $\alpha$-enhanced distribution. For a given $Fe$ content, 
the latter distribution has a larger CNO abundance (by a factor of two) compared to the normal $\alpha$-enhanced mixture. 
We discuss whether this is consistent
with existing photometric constraints, from both $HST$ and ground based photometry.
The next section will briefly present our new model calculations, while the subject of Section~3 is the comparison with photometric data.
A discussion follows in Section~4.

\section{Stellar models}\label{models}

We consider as
representative of the \rq{normal}\rq\ stellar population in NGC~1851 models computed with the
$\alpha$-enhanced mixture ($<$[Fe/H]$>$=0.4) employed by Pietrinferni et al. (2006), 
a metal mass fraction $Z$=0.002 and $Y$=0.248, corresponding to
[Fe/H]=$-$1.31 (Yong \& Grundahl~2007 have determined a mean abundance [Fe/H]=$-1.27\pm$0.03).
We have then computed additional models for
a second heavy element mixture, with abundances representative of the extreme CNONa 
anticorrelations (hereafter \lq{extreme}\rq\  mixture) detected in GCs.
This mixture is the same adopted by Salaris et al. (2006) and displays an 1.8~dex increase of the N abundance, an 0.6~dex decrease
of C, an 0.8 dex increase of Na and an 0.8 dex decrease of O, with respect our \lq{normal}\rq\ $\alpha$-enhanced heavy elements distribution.
We have taken into account the effect of this new mixture in both the radiative opacity (see Salaris et al.~2006 for more details)
and nuclear burning network, and use the same evolutionary code and physical inputs adopted
for the BaSTI stellar library (Pietrinferni et al. 2004, 2006, Cordier et al. 2007).
To ensure the same [Fe/H] in both populations, 
this set of stellar models and isochrones has ben computed by using a value of Z
equal to 0.0037 and two different He contents: Y=0.248 and 0.28. The reduction in the hydrogen abundance 
implied by this increase in Y affects the [Fe/H] value by less than 0.02~dex. 
The sum of the CNO element abundance in the isochrones with the extreme composition is a factor of 2 larger 
than in case of the normal $\alpha$-enhanced counterpart. 

Bolometric luminosities and effective temperatures of the models and isochrones for the
two metal mixtures have been transformed into magnitudes and colors using
in the transformations presented by Bedin et al~(2005), for [Fe/H]=$-$1.3. These transformations are in principle suitable
only for a scaled solar heavy element mixture. When comparisons are
performed in the $V-(V-I)$ plane, or the corresponding F606W-(F606W-F814W) plane in the ACS filter system,
the inconsistency is minimized because $(V-I)-T_{eff}$
transformations are to a good approximation independent of the
metal abundance and their distribution (e.g. Alonso et al.~1996, 1999, Cassisi et al.~2004).

\section{Comparison with the observations}

To fix the cluster distance modulus
we fit the theoretical Zero Age HB (ZAHB) of the normal population to the lower envelope of the HB in 
the M07 photometry, and adjust the reddening
by fitting the theoretical ZAHB the vertical part of the observed HB, as displayed in the upper panel of Fig.~1.
It is important to notice that M07 photometry includes only  magnitudes and colors of the RR Lyrae population
taken at random phases, therefore the portion of the observed HB crossing the instability strip cannot be used 
to put any constraint on the ZAHB fitting.

An apparent distance modulus $(m-M)_{F606W}$=15.52 together with a reddening $E(F606W-F814W)=0.038$, 
corresponding to $E(B-V)$=0.04 (according to the
relationships by Bedin et al.~2005) enables to fit the observed HB with models for both the normal population 
and the extreme one with Y=0.248. This reddening agrees with the standard estimates $E(B-V)=0.02\pm$0.02 
in the literature (see, e.g. Walker~1998)
The red part of the HB appears to be matched better by the ZAHB of the normal population,
whereas the knee at the blue side of the HB can be matched better by the ZAHB for the extreme population. Along the vertical part of the HB
the two ZAHB sequences practically overlap. On the other hand, it is difficult to accommodate the presence of an extreme population with $Y$=0.28
together with the normal $\alpha$-enhanced one. The ZAHB with $Y$=0.28 is largely overluminous, and is compatible with the observed CMD
only at the fainter blue end of the observed HB. 

In passing, we note that the ZAHB for the extreme population with Y=0.248 is mildly brighter that the ZAHB corresponding to the normal population.
This occurrence could appear at odds with the fact that usually the ZAHB locus becomes fainter when increasing the global metallicity. However, 
in this case the extreme population has a larger CNO abundance, and a larger CNO abundance implies a more efficient H-burning
shell and, in turn, a brighter ZAHB.

The top panel of Fig.~2 displays the upper MS-TO-SGB part of M07 CMD. A 10~Gyr isochrone for the normal population and a 9~Gyr isochrone
for the extreme population with Y=0.248 both match very well the bright SGB,
once shifted for the adopted distance modulus and reddening. The faint SGB is equally well matched by a 11~Gyr isochrone
for the normal population and a 10~Gyr isochrone for the extreme population with Y=0.248.
It is essentially the increased CNO abundance that affects the TO and SGB brightness in the isochrone computed with the extreme mixture, and 
causes a 1~Gyr age difference when matching either of the two SGBs.
The lower MS and the RGB of the isochrones for the two heavy element mixtures are practically coincident, thus satisfying the
constraint posed by the narrow observed MS and RGB.
We have therefore two possibilities. If the bright SGB is matched by the normal population, the two components will be essentially coeval,
with an age of 10~Gyr. If the bright SGB is matched by the 9~Gyr extreme population with Y=0.248, the normal population has to be 2~Gyr older,
in order to match the faint SGB.
This comparison demonstrates that it is possible to reproduce the peculiar SGB morphology of NGC~1851 by considering two distinct
stellar populations, with the same initial He content: a standard $\alpha$-enhanced one,
and one with a pattern of CNONa anticorrelations and a CNO sum increased by a factor of two. 

There is still a potential ambiguity regarding the ages of the two populations, depending on which one matches the bright SGB.
In absence of direct spectroscopical measurements of CNO abundances, there is no reason to prefer any of the two solutions.
Some help comes from M07 comparison of the bright to faint SGB population ratio, with the ratio of the HB stars redder than 
the RR~Lyrae gap to the ones bluer than the gap. Taking into account the fact that, as discovered by previous studies,
the RR Lyrae population in the cluster is $<$10\%  of the total HB population, the red to blue HB star ratio is close to the
55/45 ratio of bright to faint SGB populations. Based on this similarity, together with the
fact that the ZAHB for the extreme composition (Y=0.25) fits better
the blue side of the HB (see Fig.~1 and previous discussion), it is tempting to conclude that the extreme population corresponds to the faint SGB, and
therefore it is coeval with the normal $\alpha$-enhanced population, both having an age of about 10~Gyr.
We note also that this scenario in which the extreme population with Y=0.25 populates the blue side of the HB distribution, requires a 
a more efficient mass-loss during the RGB stage, compared to the normal $\alpha$-enhanced population. 
The lower panel of Fig.~1 displays the value of the ZAHB mass as a function of the color (F606W-F814W), corrected for NGC1851 reddening. 
To be located at the blue side of the RR Lyrae gap ($(m_{F606W}-m_{F814W})<$0.2) the extreme population needs to reach masses lower than   
$\sim 0.58M_{\odot}$. On the other hand, a normal population at the red side of the gap ($(m_{F606W}-m_{F814W})>$0.5) needs  
ZAHB masses larger than $\sim 0.62M_{\odot}$. Given that the TO mass for the two populations and the common age of 10~Gyr is the same 
within 0.003 $M_{\odot}$, different amount of mass have to be lost along the RGB by the two populations. We also notice that, 
for a fixed value of the $\eta$ parameter in the Reimers~(1975) mass loss formula, according to our calculations, the two distinct stellar populations 
would be expected to have approximately the same total stellar mass at the RGB tip. 
We postpone a more detailed investigation of this issue to a forthcoming paper.

The lower panel of Fig.~2 displays a fit to the SGB region using the helium enhanced extreme isochrones and the same distance modulus
as before. Brighter and fainter SGB can be matched with ages essentially equal to the case of $Y$=0.248. Due to the higher He content,
the displacement between the extreme and normal isochrone along both the MS and the RGB
is of the order of $(m_{F606W}-m_{F814W})=0.02$ mag, still marginally consistent with the
spread in M07 photometry. However, the two populations cannot coexist along the HB, unless the He enhanced one is located at the extreme end of the
blue HB, where there are very few objects. In this case the number ratios between the two populations along the HB
would not match the SGB number ratios, for any of the two possible age combinations. 

Irrespective of the choice of Y, along the RGB the combination of the normal and extreme population 
produces a well defined bump in the luminosity function (see Salaris et al.~2006) whose brightness is only marginally 
changed when considering an enhanced He abundance in the extreme component.

\section{Discussion}

In the previous section we have shown that it is possible to reproduce the peculiar SGB morphology of NGC~1851, by
considering the presence of two stellar populations with distinct initial chemical composition: a normal $\alpha$-enhanced
population and one characterized by a CNONa anticorrelation pattern with an increased CNO total abundance. 
The best agreement with the CMD and population ratios
along SGB and HB is obtained for a coeval age of 10~Gyr, and an He abundance of the extreme population showing
a negligible enhancement compared to the normal population.
It is important to notice that a consequence of the increased CNO sum in the extreme population is 
the negligible time delay between the formation of these two components, 
an occurrence consistent with current theoretical ideas about the sources of chemical pollution, 
as briefly mentioned in the Introduction. 
The peculiarity of this cluster is that there are possibly two subpopulations with well distinct element abundance ratios, not a wide range
of abundance anomalies, as observed in other clusters. Also, the negligible enhancement of He in the extreme population puts 
additional severe constraints on the models for the origin of this component.

In absence of detailed spectroscopical determinations of the CNO sum in NGC1851, 
our result cannot be a definitive proof that this interpretation of the SGB splitting is the one corresponding to
reality, just a working hypothesis that we hope will stimulate further spectroscopic work to definitely prove (or disprove) this scenario.
Yong \& Grundahl~(2007) have recently determined only oxygen abundances, and on a very small sample of stars. Our assumed metal mixtures 
imply a bimodality in the CNO abudance as a whole, but also in the individual abundances of these three elements. 
Yong \& Grundahl~(2007) sample of 8 stars stars is too small to determine with some significance whether the measured O abundances are bimodal 
or not. 
We conclude by mentioning an independent empirical result that may provide some additional support to our idea, coming
from the Str\"omgren photometry of the RGB of NGC1851, presented in Calamida et al.~(2007 - hereafter C07).
After a careful selection of the cluster members, C07 have shown that the RGB of NGC~1851
splits into two distinct and well separated sequences in the Str\"omgren $(m_1, u-y)$ diagram. At $(u-y)\sim 3.0$ the difference in $m_1$
between the two sequences is of about 0.1~mag. From the empirical relationships between $m_1$, $(u-y)$ and [Fe/H] presented in C07, one
derives that a [Fe/H] difference by more than $\sim$0.6~dex is necessary to produce this split.
The observed narrow MS and RGB in the ACS CMD and the results by Yong \& Grundahl~(2007) 
do not support this interpretation. An alternative explanation is the presence of two 
separate populations with different CN ratios given that, as discussed 
by, i.e. Bell \& Gustafsson (1978) and Dickens, Bell \& Gustafsson~(1979), variations in CN abundances
affect strongly the $m_1$ index.

\acknowledgments{}

We wish to thank F.~D'Antona, E.~Carretta and an anonymous referee, for insightful comments and remarks  
that helped improve the presentation of our results.


\begin{figure}[t]
\plotone{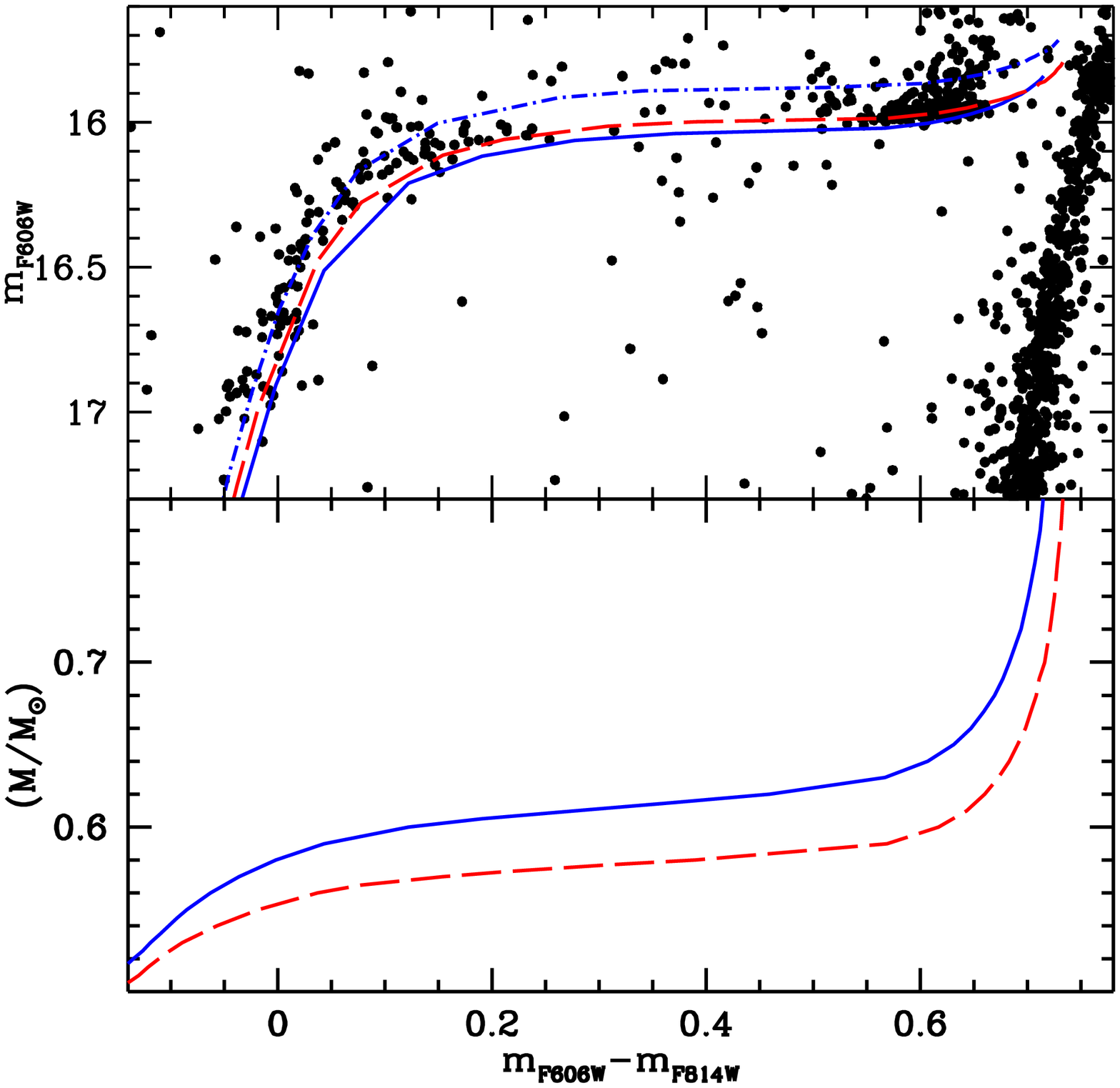}
\caption{Upper panel: $m_{F606W}$ {\sl versus} $(m_{F606W}-m_{F814W})$ diagram of NGC~1851 HB stars, from M07. Overimposed are ZAHB models for the
normal population (solid line), and the extreme population with, respectively  Y=0.248 (dashed line) and Y=0.28 (dot-dashed line) shifted by
$(m-M)_{F606W}$=15.52  and $E(B-V)$=0.04 (see text for details). Lower panel: relationship between the stellar mass and 
the $(m_{F606W}-m_{F814W})$ color for ZAHB models. The meaning of the different lines is as in the upper panel.}
\label{fig1}
\end{figure}

\begin{figure}[t]
\plotone{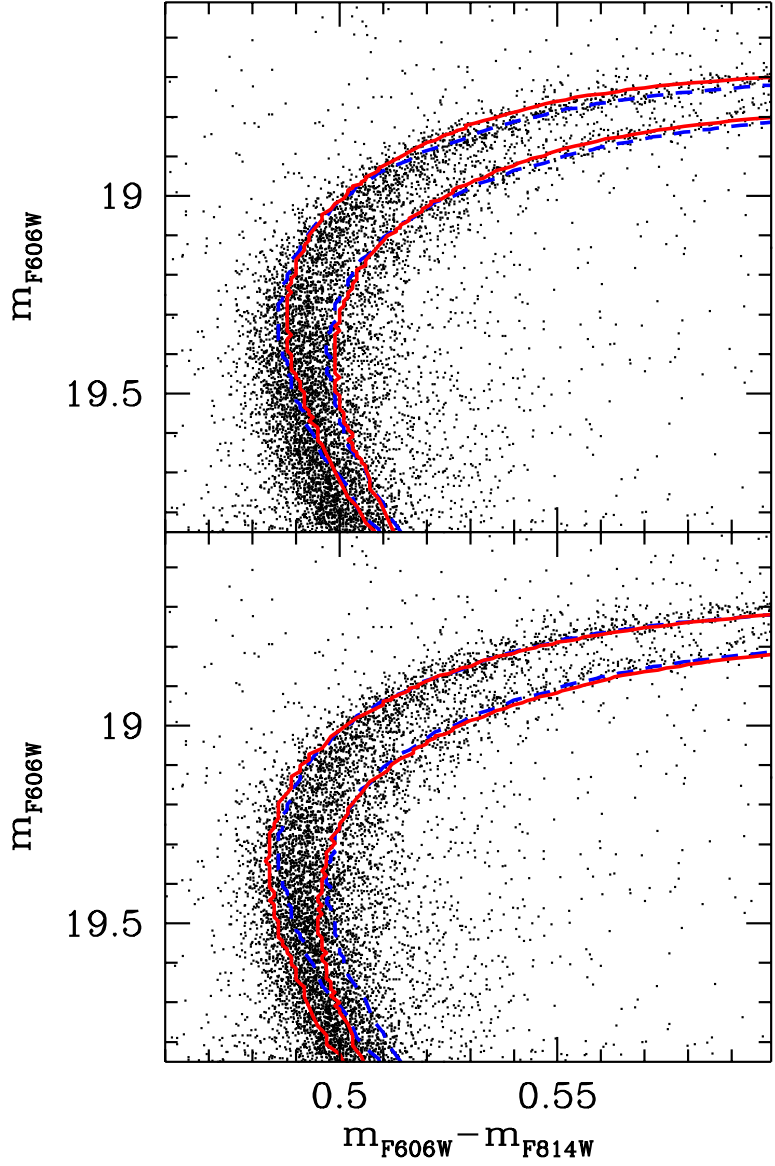}
\caption{Upper panel: $m_{F606W}$ {\sl versus} $(m_{F606W}-m_{F814W})$ diagram of NGC~1851 by M07 zoomed around the SGB. The solid lines
represent the isochrones for the extreme population with Y=0.248, and ages of 9 and 10~Gyr, respectively. The dashed lines
are isochrones for the normal population and ages of 10 and 11~Gyr. The isochrones are shifted by $(m-M)_{F606W}$=15.52
and $E(B-V)$=0.04 (see text for details).
Lower panel: as above, but this time the solid lines represent the extreme population with Y=0.280. The ages are again 9 and 10~Gyr.}
\label{fig2}
\end{figure}

\end{document}